\documentclass[prl, twocolumn]{revtex4-2}
\usepackage{amsmath,amsfonts}
\usepackage{graphicx}

\begin{document}
\title{Non-trivial Lyapunov spectrum from fractal quantum cellular automata. }
\author{David Berenstein, Brian Kent}

\affiliation{Department of Physics, University of California, Santa Barbara, CA 93106}

\begin{abstract}
A generalized set of Clifford cellular automata, which includes all Clifford cellular automata, result from the quantization of a lattice system where on each site of the lattice one has a $2k$-dimensional torus phase space. The dynamics is a linear map in the torus variables and it is also local: the evolution depends only on variables in some region around the original lattice site. Moreover it preserves the symplectic structure. These are classified by $2k\times 2k$ matrices with entries in Laurent polynomials with integer coefficients in a set of additional formal variables.
These can lead to fractal behavior in the evolution of the generators of the quantum algebra. Fractal behavior leads to non-trivial Lyapunov exponents of the original linear dynamical system. The proof uses Fourier analysis on the characteristic polynomial of these matrices.
\end{abstract}

\maketitle

Generic classical dynamical systems are chaotic. Given initial conditions are near each other, their differences grow exponentially. The Lyapunov exponents  measure these growing deviations. If we have
a system with $n$ degrees of freedom (more precisely, a dynamical system on an n-dimensional manifold), there are $n$ non trivial such exponents. If there is at least one exponentially growing direction, the system is chaotic (we refer the reader to \cite{Zas} chapter 4, for more information).  Recent advances in quantum theory have placed bounds on Lyapunov exponents in terms of the temperature \cite{Maldacena:2015waa} using out of time correlation functions (OTOC).  The bound can also be proved using the eigenstate thermalization hypothesis \cite{Murthy:2019fgs}.  It is important to understand the relation between Lyapunov exponents and quantum dynamics in general setups that do not involve the temperature.

Recent advances in quantum control also make it in principle possible to study the dynamics of qubit systems by direct simulations on a quantum computer. Such simulations usually involve a discrete time dynamics. If we consider systems that are also translation invariant, we  land on systems that are called quantum cellular automata (see \cite{Farrelly:2019zds} for a review). Most of these can not be understood in detail analytically. 

Usually Lyapunov exponents have to be measured numerically.
There are
a few systems where  one can understand the Lyapunov exponents analytically. Two standard examples are geodesic flows in constant negative curvature geometries, and the Arnold cat map.
We are most interested in generalizations of the Arnold cat map. The Arnold cat map is associated to the linear dynamics on the 2-torus given by 
\begin{equation}
\begin{pmatrix} x_n\\y_n \end{pmatrix} = \begin{pmatrix} 2&1\\
1&1 
\end{pmatrix} \begin{pmatrix} x_{n-1}\\y_{n-1} \end{pmatrix} \mod(2\pi).\label{eq:cat}
\end{equation}
here the $\mod(2\pi)$ indicates that the left hand side variables are each on a circle: $x\equiv x+2\pi, y\equiv y+2\pi$, giving a 2-torus.
The map is essentially linear and the Lyapunov exponents are given by the logarithm of the eigenvalues of the $2\times2$ matrix appearing in the dynamics. In other linear setups, Lyapunov exponents control the evolution of entanglement entropy between subsystems (see for example \cite{Asplund:2015osa}).

Some celullar automata that are easier to understand, because their dynamics can be simulated 
easily with a classical computer \cite{Gottesman:1998hu}. These are Clifford cellular automata and some of their generalizations.
The space of periodic Clifford cellular automata on a one dimensional lattice with periodicity one  on the lattice (and some of their generalizations) are classified by a set of $2\times 2$ matrices of determinant one \cite{schlingemann2008structure,gutschow2010time}, given by
\begin{equation}
    M= \begin{pmatrix} a(q, q^{-1}) &b(q, q^{-1})\\
    c(q, q^{-1})&d(q, q^{-1})
    \end{pmatrix}\label{eq:class}
\end{equation}
where the $a,b,c,d$ are Laurent polynomials with integer coefficients modulo $N$ for some integer $N$ and $q$ is a formal variable of translations.
These are palindromes in $q$, namely, $a(q,q^{-1})=a(q^{-1},q)$. In the case of Clifford cellular automata, we have that $N=2$. When the dynamics is iterated, we end up with the matrix of $M^t \mod(N)$. 
These dynamical systems produce three classes of behaviors: periodic dynamics, gliders and fractal behavior. Periodic dynamics occurs when the trace of $M$ is independent of $q$. Gliders occur when $\hbox{Tr}(M)$ is of type $\pm  (q^n+q^{-n})$, so that the eigenvalues of $M$ are $\pm  q^n$ and $\pm q^{-n}$. In all other cases the powers of $M$ give rise to fractals in the evolution of the operator spectrum.

In this paper we find a connection between  generalized cat maps, where there is a lattice of variables like $x,y$ above,
and also more general periodic cellular automata where there are $k$ such variables on each site (one can say the period is $k$ rather than 1). The cellular automata are a quantization of the generalized cat map. We prove that if the analog of $M$ has fractal behavior, then the original matrix that generalizes \eqref{eq:cat} has some non-zero Lyapunov exponent. The fractal behavior can be said to be a remnant of chaos in the original classical system. 

The connection asks us to start with generalizations of \eqref{eq:class} with coefficients in the integers, which preserve a symplectic form. To quantize, we take the coefficients modulo $N$. The Lyapunov spectrum  arises by computing eigenvalues of $M$ with a replacement of the $q$ variable by a pure phase $q\to \exp(i\theta)$. We show that if there is fractal behavior in the reduction modulo $N$, then there are some values of $\theta$ for which the Lyapunov exponents are not zero.

\section{Generalized cat maps and symplectic cellular automata}

A symplectic torus is a two dimensional phase space, with  periodic variables $x,y$ and a non-trivial Poisson bracket
$
\{x, y\}=1.
$ 
The periodicity is given by $x\equiv x+2\pi$, $y\equiv y+2\pi$. A cat map is a linear morphism of the $x,y$ system variables (which we call $M$) that preserves the periodicity and the Poisson bracket structure. This becomes a dynamical system when iterated.
The periodicity forces the matrix elements of $M$ to have integer values
\begin{equation}
    \begin{pmatrix}
    x(t)\\ y(t)
    \end{pmatrix}= M^t \begin{pmatrix}
    x\\ y
    \end{pmatrix}= \begin{pmatrix}
    a& b\\
    c& d
    \end{pmatrix}^ t \begin{pmatrix}
    x\\ y
    \end{pmatrix}
\end{equation}
Preserving the bracket requires that 
\begin{equation}
    M \cdot \Omega \cdot M^T = \Omega\label{eq:symp}
\end{equation}
where $\Omega$ is the symplectic matrix of the system
\begin{equation}
    \Omega= \begin{pmatrix} 0&1\\
    -1 &0
    \end{pmatrix}
\end{equation}
It is easy to show that this leads to $\det( M)=1$. Hence, the set of possible cat maps can be characterized as an element of $SL(2,{\mathbb Z})$.
For us, a generalized cat map is a similar map on $n$ copies of the symplectic torus, see \cite{Berenstein:2015yxu}. 
The linear space of periodic $x,y$ variables get converted to vectors of periodic variables $\vec x, \vec y$. 
The linear space can be described as a tensor product
$
   \hbox{Span} (\vec x, \vec y) \simeq  (x,y) \otimes {\mathbb R}^n$
and the symplectic structure becomes $\Omega\to\Omega \otimes 1$ in this setup, where $1$ is the $n\times n$ identity matrix. To have a generalized cat map, we get a matrix with $2n\times 2n$ entries, with integer coefficients such that \eqref{eq:symp} is true, where the right hand side uses the $2n\times 2n$ version of $\Omega$. If $\lambda_i$ are the eigenvalues of $M$, the Lyapunov exponents of the dynamics of $M$ are given by
\begin{equation}
    \kappa_i = \log( |\lambda_i|)
\end{equation}
The eigenvalues have the following properties. If $\lambda$ is an eigenvalue, then so is, $\bar \lambda, \lambda^{-1}, \bar \lambda^{-1}$
(see for example \cite{Asplund:2015osa}).
If $\lambda$ is a generic complex number, they come in families of $4$ eigenvalues and some $\kappa\neq 0$. If $\lambda $ is real, then there is only one paired eigenvalue $\lambda^{-1}$ and $\kappa\neq 0$. Only if $\lambda$ is unitary (and so is $\lambda^{-1}=\bar\lambda$), we get that the corresponding $\kappa$ vanish. 

{ \bf Quantization:}  Since the variables $x,y$ on a single torus are not single valued, they can not correspond to quantum observables upon quantization.
A general observable on a torus can be decomposed in Fourier series, so $\exp(i x), \exp(i y)$ generate the algebra of observables. As the torus is a compact phase space, on quantization we expect that the Hilbert space ${\cal H}$ associated to the torus is finite dimensional, $\dim {\cal H}= N$. 
Consider the quantum variables $U=\exp(i x)$, $V= \exp(i y)$. It is easy to show by the Baker-Cambell-Haussdorf  formula that $UV, VU$ differ by a phase
\begin{equation}
UV = VU \exp( i\zeta)
\end{equation}
here the phase $\zeta$ plays the role of $\hbar$, and $U,V$ are unitary operators. In a Hilbert space of dimension $N$, the $\exp(i\zeta)$ is taken as a primitive N-th root of unity. We also require $U^N=V^N=1$, so that $U,V$ are realized by  clock-shift matrices (monomials in the $U,V$ are called Weyl operators). In these circumstances we would have that  we can interpret this quantum condition as $N x\equiv 0$, so the entries of matrix $M$ are only well defined modulo $N$.  When $N$ is two, $U,V$ are Pauli matrices, and a quantum torus gets quantized into the Hilbert space of a single qubit. 
For the $n$ torus system, we have a $U,V$ pair per each torus, and the $U,V$ commute with each other if they belong to different tori. A quantized generalized cat map takes monomials in the $U_i,V_i$ variables to monomials in these variables. Because it preserves the phase space structure, it is realized as a unitary operator on the Hilbert space, and can be described as an algebra automorphism of the algebra generated by the $U,V$. When we quantize $n$ copies of the system, we take the Hilbert space to be ${\cal H}_{tot} \simeq {\cal H}^{\otimes n}$. In these systems, the algebraic properties of the Unitary makes it easy to describe the dynamics in the Heisenberg picture, where it can be simulated with classical physics \cite{Gottesman:1998hu}. This reduces to essentially computing the powers of the  matrix $M$, modulo the integer $N$ \cite{schlingemann2008structure}, which is usually taken to be prime (non-primes lead to the Clifford group not being a unitary 2-design \cite{PhysRevX.8.021014}).

\section{Fractal cellular automata }

We are now ready to start working on the problem of cellular automata. We will require that the dynamics of the system take place 
on a lattice. For simplicity we will choose a one dimensional lattice first, and then we will explain how to go to higher dimensions. 
At each site a cell will contain $k$ (quantum) tori. We want a dynamics that is translation invariant on the lattice and that is also 
a generalized cat map. The variables are therefore classified by a set
$
    x_{i, \alpha}, y_{i,\alpha} $
where the $i\in\{1, \dots k\}$ and $\alpha \in {\mathbb Z}$ is the lattice site location. Let us also call the set of all the $x_{i,\alpha},y_{i,\alpha}$ at a single lattice site by $z_{I,\alpha}$, and $I\in \{1, \dots 2k\}$.
There is a translation operator on the lattice that we will label with the variable $q$, which acts as follows
\begin{equation}
q( x_{i, \alpha}) = x_{i,\alpha+1}, \quad q(y_{i, \alpha})= y_{i,\alpha+1}.
\end{equation}
It is easy to check that $q$ is a matrix similar to $M$ with integer coefficients. Also, the transpose of $q$ is $q^T= q^{-1}$.
A periodic dynamics of the $\vec x, \vec y$ variables is a map of the form
\begin{eqnarray}
    z_{I,\alpha}(t) &=& \sum_{J,\beta}M_{IJ,\alpha\beta}(t) z_{J, \beta}
\end{eqnarray}
where the $M$ are finite integers, and only finitely many are non-zero. This is a locality property.
It is periodic if $M_{IJ, (\alpha+k)(\beta+k)}=M_{IJ,\alpha\beta}$. We can use the images of the $z_{I,0}$ to get the full dynamics.
For example, $z_{I,s}= q^s(z_{I,0})$. This is also true if we use the time translates $z_{I,s}(t)$.
This way we get that the image of the zero-th cell variables can be written as
\begin{equation}
    z_{I,0} (t)= \sum_{s=-\infty}^\infty  M_{IJ,s}(t) q^s(z_{J,0}).
\end{equation}
If we apply $q^r$ on the left, it is easy to see that it commutes with the action of $M$, so we can think of $M$ being defined identically by
$$
M(t)\equiv \sum_{s=-\infty}^\infty M_{IJ,s}(t) q^s =M(1)^t .\label{eq:dyn}
$$
In this presentation, $q$ can be treated as a formal variable and in the time evolution we take the $t$-th power of the matrix $M(1)$ including the powers of the variable $q$. 
We get for each  $t$,  a $2k\times 2k$ matrix of Laurent polynomials in $q$ with integer coefficients. This is denoted by saying that  $M \in M_{k\times k}[{\mathbb Z}[q, q^{-1}]]$.  It is easy to convince oneself that the symplectic condition is identical to equation \eqref{eq:symp},
where $\Omega$ is a $2k \times 2k$ symplectic matrix which is independent of $q$, whereas $M$ is q-dependent and the transpose must include $q\to q^{-1}$ as follows
\begin{equation}
M^T_{IJ}= \sum_{s=-\infty}^\infty M_{JI,s}(t) q^{-s} 
\end{equation}
If the lattice is finite and periodic, we can also impose the constraint $q^L=1$. If we think of $M$ as a $2k\times 2k$ matrix, it satisfies its characteristic polynomial of degree $2k$. The coefficients of the characteristic polynomial will also be Laurent polynomials in $q$. The symplectic condition can also be written as
$$
M^{-1}= \Omega \cdot M ^T \cdot (\Omega^{-1})
$$
which shows that the inverse dynamics is also an allowed dynamics an that the generalized Clifford quantum cellular automata have a group structure.

When we include the naive quantization condition $N z_{I,\alpha}\equiv 0$, the coefficients of $M$ are well defined only modulo $N$ and the same is true for the coefficients of the characteristic polynomial. Given $M$, we can say its naive quantization just takes $M$ with integer coefficients to their reduction modulo $N$. 
The dynamics of the $U,V$ operators is well defined given $M$ and $N$, up to a cocycle condition that tells us the phase of the $U,V$ images. We will assume that this is given 
and then the dynamics associated to $M$ becomes a quantum cellular automaton. In each cell we have a local algebra associated to the Hilbert space $({\mathbb C}^N)^k$, which generalizes the notion of a qubit algebra. The condition that monomials in the $U,V$  variables go to monomials in $U,V$ variables generalizes the notion of the Clifford cellular automata. This is also a case where the classical dynamics associated to the matrix $M$ characterizes the Heisenberg evolution of the $U,V$ operators. This is important for us now.

{\bf Self-similarity:} 
We now want to argue that the quantum dynamics associated to $M$ is self-similar and leads to fractal behavior in the iterated images of the $U,V$ variables. This is understood in the classification of Clifford cellular automata. In this more general case, the simplest way to do this is to assume that $N$ is a prime number $p$ for simplicity. Since the matrix $M$ satisfies its characteristic polynomial equation, we have that 
\begin{equation}
    M^{2k}+\sum_{0<s<2k} b_s M^s+b_0=0\label{eq:char}
\end{equation}
where $b_0=\det(M)$ and $b_s\in {\mathbb Z}_p[q, q^{-1}]$. We now want to replace $M$ by a formal  commutative variable $\xi$, which also satisfies the same polynomial equation of $M$,  where the coefficients are in the set of Laurent polynomials with coefficients in ${\mathbb Z}_p$. In this sense $\xi$ is a formal eigenvalue of the matrix $M$.
This replacements takes us from problems of matrices to a problem involving roots of polynomials with coefficients in the ring of Laurent polynomials with coefficients in ${\mathbb Z}_p$. This replacement has taken the problem from matrices to algebraic number theory. The condition that the coefficients of the Laurent polynomials are numbers modulo $p$ is expressed by saying that we are working with a commutative algebra in characteristic $p$. In this setup, there is an algebraic endomorphism of these algebras to themselves called the Frobenius endomorphism. It takes $u \to u^p$. One can easily show that for two commutative variables with coefficients in ${\mathbb Z}_p$ that $(a+b)^p = a^p+b^p$, so this produces an algebra morphism that takes products to products and is linear in sums.  When applied to equation \eqref{eq:char}, the characteristic equation of the iterates of this morphism has coefficients $(b_s)^{p^r}$. This morphism acts by taking $q\to q^p$, but otherwise leaving the coefficients of each Laurent polynomial intact. This way the iterates of powers of $p$ (iterates of the Frobenius morphism), $p^s$ for each $M$ have the same number of non-zero coefficients in each Laurent polynomial. Each time we apply this morphism, we rescale time by a factor of $p$, so we see that a simple property of the coefficients of the characteristic polynomial are self-similar. By this, we mean that we count the number of non-zero entries in the rescaled images of the map by a factor of $p$, we get a constant number. A formal proof of self-similarity can be found in \cite{GNW10,GNW10a}

We will say that the self-similar behavior has fractal dimension $d$ if the number of non-zero elements of one of the coefficients of the characteristic polynomial $b_s$ scale on average like $t^d$ at time $t$. Basically, at time $t$, we define a function $f(t)$ such that $f(t)$ is the number of non-zero coefficients of the $b_s$ Laurent polynomials. More precisely, a notion of the fractal dimension $d$ is defined by requiring that
\begin{equation}
g(s)= \sum_{t=0}^\infty  f(t) t^{-s-1}
\end{equation}
is convergent if $s>d$ and divergent otherwise. If the fractal dimension is zero, then $f(t)$ is bounded above: there is no growth in the number of non-zero coefficients $b_s(t)$. This notion of dimension seems to coincide with the fractal dimension of the cellular automaton in simple cases for one of the coefficients. We say we have fractal behavior if $d>0$. The $d=0$ case arises in the case where the celullar automaton has gliders. 
This guarantees the growth property of $f(t)$ that we need, by comparing to the harmonic series.
This is all we are going to need in what follows.

\section{Fractal behavior implies non-zero Lyapunov exponents}

Our goal is now to obtain the eigenvalues of the matrix $M$, which is a matrix with Laurent polynomials in $q$ over the integers, and we assume that the reduction modulo $N$ of $M$ has fractal dimension $d>0$. Once we decide that the matrix $M$ is over the integers, we can think of it also as a complex matrix. 

Remember that $q$ is a translation operator. The dynamics is translation invariant. This is stated by saying that $M$ commutes with $q$.
Basically, $M$ and $q$ can be diagonalized simultaneously. This is  Bloch's theorem applied to $q,M$. 
The action of $q$ is unitary, so the eigenvalues of $q$ are of the form $\exp(i\theta)$. The problem of the eigenvalues of the big matrix $M_{IJ, \alpha\beta}$ gets reduced to computing the eigenvalues of the small $2k\times 2k$ matrix $M_{IJ}(q\rightarrow \exp(i\theta))$ for each $\theta$.
If for some $\theta$ we have at least one eigenvalue of $M(q=\exp(i\theta))$ which is not in the unit circle, then we have a non-zero Lyapunov exponent. This is what we want to show. We will do so by contradiction assuming both that the reduction of $M$ modulo $N$ is fractal and that the eigenvalues of $M$ all lie in the unit circle. 

The basic idea is the following: the coefficients of the characteristic polynomial of some power of $M$, $M^t$ (which depends on $q$) are sums of $t$-powers of products of eigenvalues of $M$ evaluated at $q=\exp(i\theta)$. If two matrices have the same characteristic polynomials, then they will have the same spectrum of eigenvalues and hence the same Lyapunov spectrum. This suggests some weak notion of equivalence of the two dynamical systems.

More precisely, $b_s(t,q=\exp(i\theta))$ is a sum of products of $s$ eigenvalues. There are $N_s={2k \choose s}$ distinct products of eigenvalues $\lambda_i^t$. Each of these products is a unitary number if the eigenvalues are all unitary. The sum is bounded above  $|b_s(t,q=\exp(i\theta))|\leq N_s$. We will use this inequality by squaring it and averaging over the phase $q=\exp(i\theta)$ as follows
\begin{equation}
    \frac 1{2\pi} \int_0^{2\pi} |b_s(t,q=\exp(i\theta))|^2 d\theta \leq N_s^2 \label{eq:average}
\end{equation}
If this inequality is violated, then the hypothesis that the eigenvalues are in the unit circle for all $\theta$ is wrong, and there is some $\theta$ for which this is not true. At that value, at least one of the eigenvalues is outside the unit circle.

Let us now find a lower bound on the integral. Here we use the fact that for some $s$ and some $t$, the number of non-zero coefficients of $b_s(t)$ modulo $N$ can be as large as we need, of order $t^d$. This number can be chosen to be larger than $N_s^2$.
This implies that the polynomials of $q$ in the characteristic polynomial of $M$ has at least as many non-zero integer coefficients. After all, if a coefficient is non zero modulo $N$, it comes from a non zero coefficient in the original polynomial with integer coefficients: the characteristic polynomial reduced mod $N$ is the characteristic polynomial of $M$ reduced modulo $N$.
Each of those non-zero coefficients in the characteristic polynomial coefficient can be thought of as a Fourier coefficient of the function $b_s(t, \theta)$.
The integral \eqref{eq:average} is the $L^2$ squared norm of $b_s(t, \theta)$ and is also given by the sum of the squares of said Fourier coefficients. These are integers, so each such coefficient is greater than or equal to one in absolute value. Our lower bound on the number of non-zero entries implies that for that particular times $t$ and $s$, the bound is violated. 
This proves our main claim.

When the lattice is finite size, the same argument works so long as the lattice is large enough, and instead of summing over all $\theta$, we sum over the $L$ roots of unity in $q$. This would use a discrete Fourier transform of $b_s$.
If the lattice is bigger than one dimensional, we have a site lattice that must cover a unit cell, and we have as many $q$ translation variables as the dimension of the lattice (see for example \cite{JMP621}). These all commute with each other, so  the Bloch argument works the same. Instead of doing a single integral over a circle,  we average over a torus whose dimension is the dimension of the lattice.

\begin{acknowledgments}
We would like to thank S.  Mukherjee for discussions.  Work of D.B. supported in part by the Department of Energy under grant  DE-SC0019139.
\end{acknowledgments}

\pagebreak

\section{\large \underline{Supplementary material}}

This material complements certain observations made in the paper with illustrative examples.
The two observations that are important are the following: the characteristic equation of the 
matrix that determines the generalized Clifford cellular automata dynamical system encodes the fractal structure of the cellular automata and therefore it is possible to extract the fractal dimension from it.

\section{Fractal in the characteristic equation}

We will deal with two standard cellular automata. The simplest Clifford cellular automata of period one  and  the CNOT gate system studied in \cite{Berenstein:2021wzq}.

For the first system, studied in \cite{}, the $2\times 2$  matrix determining the dynamics is given by 
\begin{equation}
M_{2\times 2}= \begin{pmatrix} 1+q+q^{-1} &1\\
1&0
\end{pmatrix}
\end{equation}
The characteristic equation for $M$ is 
\begin{equation}
    \xi^2 - \hbox{tr}(M) \xi +1=0
\end{equation}
For the iterates of $M$, $M^n$, the characteristic equation is determined entirely by $t_n=\hbox{tr}(M^n)$. Since $M$ satisfies it's own characteristic equation, and if we multiply by $\xi^{n-2}$ and take traces, we get the recursion
\begin{equation}
    t_n- t_1 t_{n-1}+t_{n_2}=0\label{eq:rec1}
\end{equation}
with initial conditions $t_0=0$ (the trace of the $2\times2$ identity matrix is zero modulo 2), and $t_1=1+q+q^{-1}$.

\begin{figure}[ht]
\includegraphics[width=13 cm]{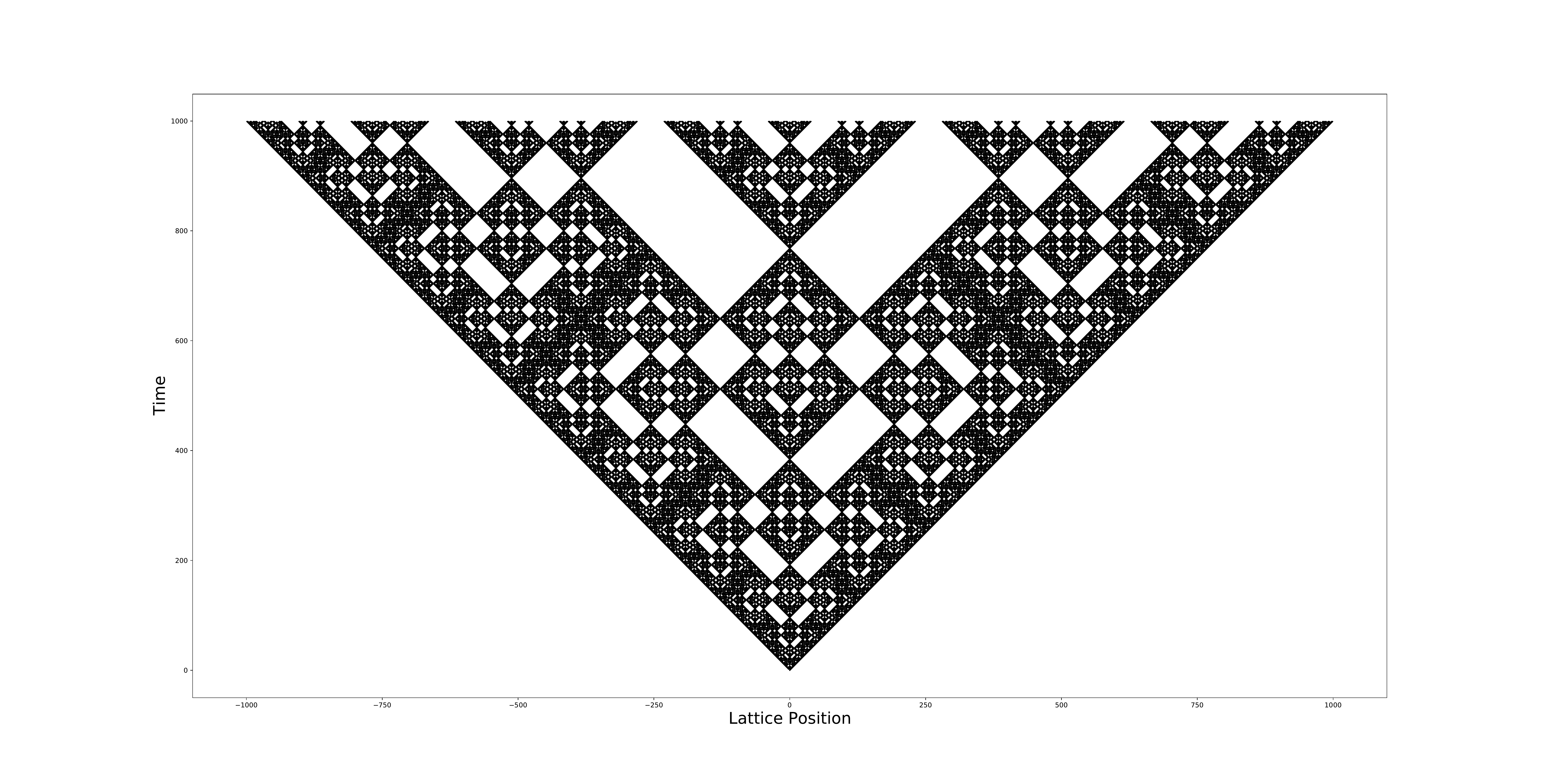}\caption{Fractal for the traces of the standard quantum cellular automaton with period 1}\label{fig:frac1}
\end{figure}

The polynomial $t_n \simeq q^{-n}+\dots + q^n$ is of order $n$. To see the fractal, we count the positions $a_{k,n}\neq 0$ in the polynomial. We get the picture in figure \ref{fig:frac1}

A fit to the dimension of the fractal shows that the fractal dimension is $h\simeq 1.830$. This implies that the number of terms at each time grows
like $t^0.83$.

\begin{figure}[ht]
\includegraphics[width=13 cm]{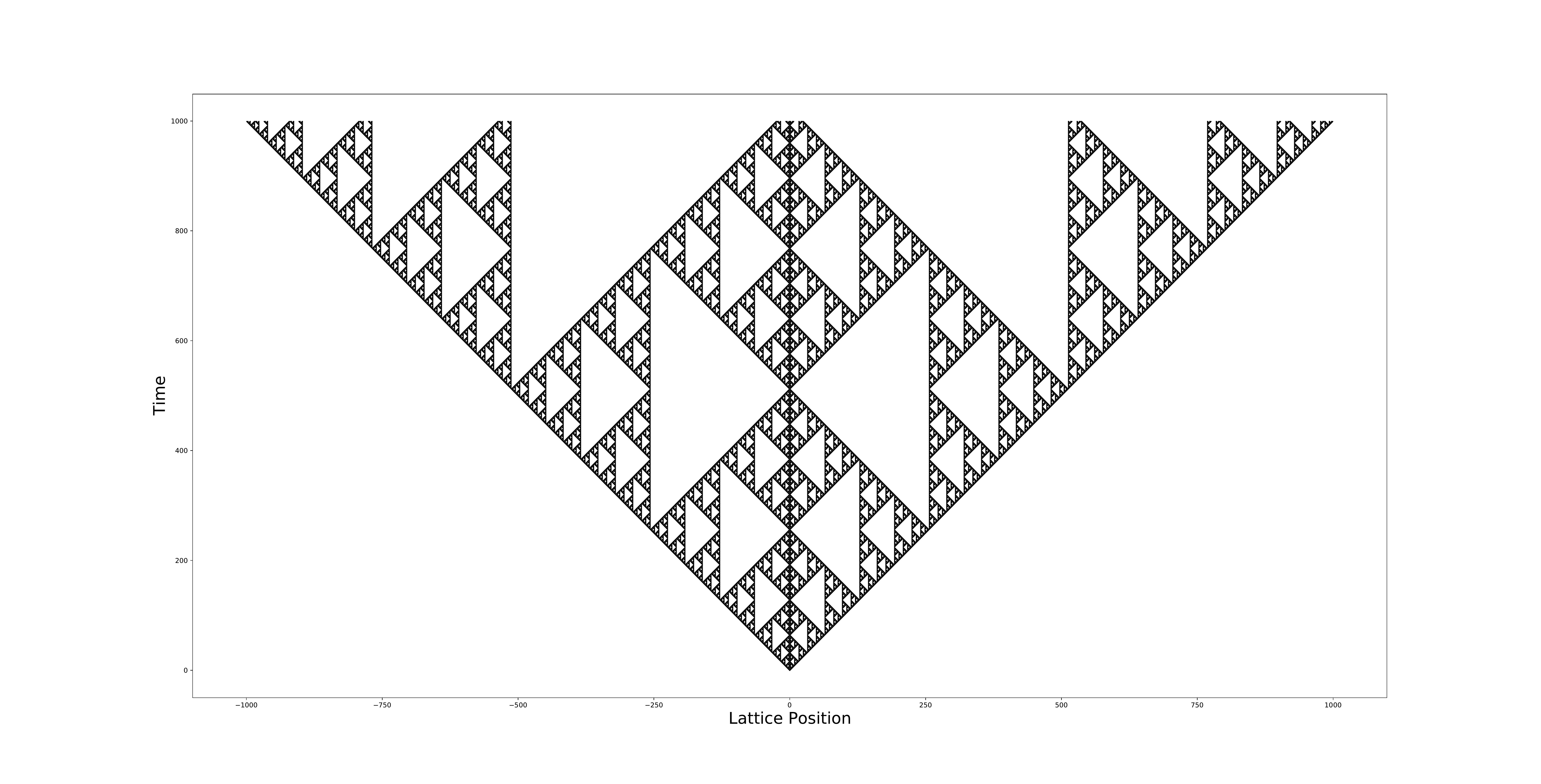}\caption{Fractal for the traces of the CNOT gate}\label{fig:sier}
\end{figure}

For the second system, the CNOT cellular automata, the matrix is given by
\begin{equation}
    M_{4\times 4}=\begin{pmatrix} q+1&q&0&0\\
    1&1&0&0\\
    0&0&1&1\\
    0&0&q^{-1}&1+q^{-1}
    \end{pmatrix}
\end{equation}
where we also have that the symplectic form is
\begin{equation}
    \Omega= \begin{pmatrix} 0& 0&1&0\\
    0&0&0&1\\
    -1&0&0&0\\
    0&-1&0&0
    \end{pmatrix}
\end{equation}
As can be seen, the matrix $M$ splits into two blocks, each a $2\times 2$ matrix. This persists when we take powers. Call the matrices
\begin{eqnarray}
    M_1&=& \begin{pmatrix}q+1&q\\1&1\end{pmatrix}\\
    M_2&=& \begin{pmatrix} 1&1\\q^{-1}&q^{-1}+1\end{pmatrix}
\end{eqnarray}
and set $\omega_n=\hbox{tr}(M_1^n)$, $\tau_n=\hbox{tr}(M_2^n)$. The characteristic equation of $M^n$ is 
\begin{equation}
(\xi^2-\omega_n\xi +1) (\xi^2-\tau_n \xi +1)   =0
\end{equation}
and it factorizes. 
The trace of $M^n$ is equal to $\tau_n+\omega_n$ and each of these satisfy an independent recursion like \eqref{eq:rec1}. This sum is also a palindrome. This is where we find the Sierpinski triangle fractal in the characteristic equation: the figure is a double copy, see the figure
\ref{fig:sier}. 
A fit to the dimension shows that $h\simeq 1.585$, so the number of terms at each time grows as $t^{0.530}$.

\end{document}